\input amstex.tex
\documentstyle{amsppt}
%\magnification\magstep1
\topmatter
\title CURRENT ALGEBRA AND EXOTIC STATISTICS IN 6 DIMENSIONS \endtitle
\author Jouko Mickelsson \endauthor
\affil Theoretical Physics, Royal Institute of Technology, S-10044,
Stockholm, Sweden. e-mail: jouko\@theophys.kth.se \endaffil
\endtopmatter

\document
\NoBlackBoxes
\baselineskip=18pt

%\redefine\Bbb{\bold}

\define\a{\alpha}

\redefine\b{\beta}

\define\gm{\bold g}

\define\<#1,#2>{\langle #1,#2\rangle}
\define\dep(#1,#2){\text{det}_{#1}#2}
\define\norm(#1,#2){\parallel #1\parallel_{#2}}

\define\TR{\text{tr}\,}
ABSTRACT By studying ordinary chiral fermions in background gauge fields we show
that in the case of gauge group $SU(3)$ and space-time dimension $5+1$
localized solitons obey $q-$ commutation relations with $q$ not equal to
$\pm 1$ but a third root of unity.

\vskip 0.5in

1. INTRODUCTION

\vskip 0.3in
A topological explanation of the spin-statistics relation in quantum
field theory is an old idea which was proposed by Finkelstein and Rubinstein
in 1968, [FR]. Later, in 1984 Witten studied a concrete model (Wess-Zumino
model) in $3+1$ space-time dimensions and showed that there are solitonic
field configurations which behave like fermions under rotations. The
essential property of the model is that there is a nonlocal (in $3+1$
dimensions) term which can be written as an integral of a 5-dimensional
topological density, [W]. One can also show that an interchange of a two
nonoverlapping solitons of odd degree produces a phase factor $-1$ of
the corresponding states in Fock space. This is a consequence of the
properties of the way the gauge group acts in the Fock space, [M1].

In this paper we shall study a more exotic statistics in higher space-time
dimensions. In $3+1$ dimensions classical solitons can be quantized
only as fermions or bosons. In $1+1$ it is known that there are models
with fractional statistics. We shall show that also in $5+1$ dimensions
there is an option to quantize solitons such that they do not obey
neither canonical commutation nor anticommutation relations but
relations of the type
$$q(f_1) q(f_2) = e^{2\pi i/3} q(f_2) q(f_1).$$
This relation is obtained starting from a quantization of ordinary chiral
fermions in external gauge field with the gauge group $SU(3)$ and looking
at the states which are produced from the vacuum by the action of
topologically nontrivial gauge transformations.
As in Finkelstein's and Rubinstein's original idea, everything goes back
to a computation of certain homotopy groups. In Witten's case the solitons
were classified by a winding number corresponding to an element in $\pi_3(
SU(2))$ and the fermionic property under rotations is explained using
$\pi_4(SU(2)) =\Bbb Z_2.$ In our case we just increase dimensions; $SU(3)$
solitons in 5 dimensions are labelled by $\pi_5(SU(3)) =\Bbb Z$ and the
exotic statistics corresponds to $\pi_6(SU(3)) =\Bbb Z_6.$

\vskip 0.3in
2. ACTION OF THE GAUGE GROUP ON FERMIONIC VACUA

\vskip 0.3in
In this section we shall explain some results in [CMM1, CMM2] which are
necessary for the discussion in Section 3.

Let $M$ be the compactified physical space which is assumed to be a manifold
without boundary, of odd dimension $d=2n+1$ and with a fixed metric, spin
structure, and orientation. In order to describe the minimal coupling
of fermions to external Yang-Mills potentials we define the fermionic
one-particle Hilbert space as $H= L^2(S) \otimes V,$ where the first factor
in the tensor product is the space of square-integrable sections of the
spin bundle (with fiber dimension $2^n$) and $V$ is a complex finite
dimensional inner product space. A compact gauge group $G$ acts in $V.$
A vector potential is a smooth 1-form $A$ on $M$ with values in the
Lie algebra $\gm$ of $G.$ We denote by $\Cal A$ the affine space of all
vector potentials. Each $A\in \Cal A$ defines a Weyl-Dirac hamiltonian
$D_A$ acting on smooth sections in $H.$ Note that we are here dealing
with massless (left handed, say) fermions.

The Hilbert space $H$ splits to $H= H_+ \oplus H_-,$ where $H_+$ corresponds
to positive energies for a fixed (background) hamiltonian $D_0 =D_{A_0}$
and $H_-$ is its orthogonal complement. This polarization defines a
representation of the canonical anticommutation relations  with a Dirac
vacuum $|0>.$ The technical difficulty is that for different vector
potentials $A$ the representations of the CAR algebra are inequivalent.
Geometrically, the Fock spaces for different potentials form a smooth
vector bundle $\Cal F$ where the fibers $\Cal F_A$ carry inequivalent
CAR representations. However, the bundle can be trivialized as
$\Cal F \simeq \Cal A  \times \Cal F_0$ because the base space $\Cal A$
is affine. Except for the case $d=1$ there is no natural trivialization.

In general, the gauge action of $\Cal G=Map(M,G)$ in $\Cal A$ cannot be lifted
to the total space of the bundle $\Cal F$ such that $g^{-1} \hat D_A g =
\hat D_{A^g}$ for all $A\in\Cal A$ and $g\in\Cal G.$ Here $\hat D_A$ is the
second quantized hamiltonian and $A^g = g^{-1} A g +g^{-1}dg.$
Instead, there is a nontrivial
extension $\hat\Cal G$ which acts in $\Cal F.$ The fiber of the extension
is $Map(\Cal A, S^1).$ If $d=1$ then one can reduce the fiber to
the subgroup $S^1$ of constant functions on $\Cal A,$ defining a central
extension of the loop group $LG$ for $M=S^1.$

On the Lie algebra level, the extension of the group of gauge transformations
is given by a 2-cocycle (Schwinger term),
$$[q(X), q(Y)] = q([X,Y]) + c(X,Y;A),$$
where $q(X)$ denotes the second quantization of the infinitesimal gauge
transformation $X$ and the cocycle $c$ is a linear function of $X,Y$,
antisymmetric in the arguments $X,Y$, and satisfies
$$c(X,[Y,Z];A ) + \Cal L_X c(Y,Z;A) + \text{cycl. permutations of $X,Y,Z$}
=0.$$
Here $\Cal L_X f$ denotes the infinitesimal gauge variation of a function
$f=f(A).$ If $d=1$ then
$$c(X,Y;A) = \frac{1}{2\pi} \int \TR X dY$$
does not depend on $A.$ The trace is computed in the vector space $V.$
For $d=3$ one has, [M2, F-Sh],
$$c(X,Y;A) = \frac{i}{24\pi^2} \int \TR A [dX,dY].$$

Given a vector $\psi\in \Cal F_A$ we have a connection in the complex line
bundle $E$ over the gauge orbit $A^{\Cal G};$ the fiber at $A^g$ is spanned
by the vector $\hat g^{-1} \psi,$ where $\hat g$ is any element in the fiber in
$\hat\Cal G$ over $g.$ The curvature of the line bundle at $A$ is given
by the 2-cocycle $c.$

We shall study the case when the local anomaly vanishes, that is, the
Schwinger term $c(X,Y;A) \equiv 0.$ There still might be a global anomaly
which manifests itself as a nontrivial holonomy along gauge orbits in
vacuum subbundles of the Fock bundle. (In the case of Witten's anomaly
this was analyzed in [NA].)
As discussed in [CMM2], this leads to an extension of
$\Cal G$ by a finite center $Z.$ The center is equal to $\pi_{d+1} (G).$
Whether it acts faithfully or not in Fock spaces has to be determined
separately for each choice of $G,$ dimension $d,$ and the representation
of $G$ in $V.$ We shall concentate to the physically interesting case
$G=SU(3).$ If $d=3$ we have $\pi_4(G)=0$ and there is no global anomaly.
The first nontrivial case is $d=5$ and $\pi_6(G) = \Bbb Z_6.$
It is known that for this choice the extension $\hat G$ acts faithfully
in the Fock bundle, [CMM2]. In other words, a parallel transport  of
a vector in the Fock space around a closed loop in a gauge orbit in
$\Cal A$ produces a phase $e^{k \cdot 2\pi i/6}$ corresponding to the
element $k\in \Bbb Z_6$ defined by the loop in $Map(M,G).$

\vskip 0.3in
3. GENERALIZED STATISTICS

\vskip 0.3in
We now pose the following problem. Consider a pair of localized soliton
configurations $f_1,f_2 \in \Cal G=Map(M, G).$ According to Section 2 we can
construct quantum operators $\hat f_1$ and $\hat f_2$ acting on second
quantized fermions, or more precisely, we have operators acting on sections
of the bundle of fermionic Fock spaces parametrized by external vector
potentials. The operators $\hat f_i$ are uniquely determined up to a
phase, which in the case when $\pi_1(\Cal G)=Z_n$ and $H^2(\Cal G)=0$ can
be taken as a $n:th$ root of unity. Since the group of quantum gauge
transformations is a $Z_n$ extension of $\Cal G$ we have
$$\hat f_1 \cdot \hat f_2 = e^{i\beta_{12}}\widehat{f_1f_2}
\text{ with $n\beta/2\pi \in \Bbb Z$ }.$$
If we assume that $f_1,f_2$ have nonoverlapping support then $f_1 f_2=
f_2 f_1$ and it follows that
$$\hat f_1 \cdot \hat f_2 = e^{i\alpha} \hat f_2 \cdot \hat f_1, \text{ with
$\a=\b_{12} - \b_{21}.$ } $$
In [M1] it was proven that in the case of space-time dimension equal to
$3+1$ and $G=SU(2)$ the phase factor is equal to $-1$ if the winding numbers
of the solitons are odd; the phase is equal to $+1$ if the winding number
of at least one of the solitons is even. Thus odd solitons behave like
fermions and even solitons like bosons.

We shall next generalize the above result to higher dimensional space-times
and other gauge groups. The main result is that in higher dimensions one
can construct anyonic statistics, the phase $e^{i\alpha}$ is some root of
unity. This is contrary to the common wisdom according to which anyonic
statistics occurs only in $1+1$ and $2+1$ dimensional field theory models
and in the framework of Wightman axioms all particles are either fermions
or bosons in higher dimensions. Here we are not considering a quantization
of point particles but extended smooth solitonic objects, so there is no
contradiction with Wightman axioms.

We shall start by studying the case $d+1=5+1$ and $G=SU(3).$ The physical
space $M$ is assumed to be compactified as the sphere $S^5.$ Since $\pi_5(
SU(3))= \Bbb Z$ there are solitons in $Map(M,G)$ and these are classified,
up to homotopy, by integers.

Let $g(t),$ with $0\leq t\leq 1,$ be a path in $SU(3)$ connecting the indentity
to a nontrivial
element of the center $\Bbb Z_3 \subset SU(3).$ For $f_0\in Map(M,G)$ we
can define a function $f(t,x)$ on $S^1 \times S^5$ by
$$f(t,x)= g(t) f_0(x) g(t)^{-1}.$$
Note that $f(t,x)$ is periodic in $t$ because $g(1)$ commutes with everything.
We want to show that this map represents an element of order 3 in $\pi_6(G).$
For a discussion of the homotopy groups of classical Lie groups needed in this
paper see e.g. [DP].

First we observe that $f^3 = g(t)^3 f_0(x)^3 g(t)^{-3}$ and $g(t)^3$ is a
closed loop in $G$ because of $g(1)^3=1.$ But $SU(3)$ is simply connected
and therefore $g(t)^3$ contracts to a constant loop and so $f^3$ is homotopic
to the identity element in the group $\pi_6(G).$ It remains to show that
$f$ itself is not homotopic to a constant loop in $Map(M,G).$ This is most
conveniently done by computing the integral
$$I(h)=\left(\frac{1}{2\pi}\right)^4 \frac{1}{840} \int_B \TR (dh h^{-1})^7,$$
where $D$ is the unit disk and $h: D \times S^5 \to SU(4)$ is a map such that
at the boundary
$S^1 \times S^5$ $h(t,r=1,x) = f(t,x).$ The extension $h$ exists because
$\pi_6(SU(4)) = 0.$ The integral $I(h)$ is a homotopy invariant \it on
a closed 7-manifold. \rm On the boundary $S^1\times S^5$ the integral
defines a homotopy invariant for $f$ modulo integers. This is because 1)
the form $\TR(df f^{-1})^7$ vanishes identically on $SU(3),$ 2) gluing
together a pair of $h'$s with the same boundary values produces a map
for which the value of the integral is an integer.

The integral $I(h)$ can be evaluated by observing first that we can replace
$g(t) f_0(x) g(t)^{-1}$ by  $g_1(t) f_0(x) g_1(t)^{-1}$
where $g_1(t)$ is a closed loop in the bigger group $SU(4).$ This follows
from the fact that the path $g(t)$ is homotopic (with end points fixed) to
the path $\tilde g(t)=\exp(it X),$ where $X= diag(2\pi/3, 2\pi/3, -4\pi/3).$
Here we have chosen (from the two nontrivial elements in the center)
$z=diag(e^{2i\pi/3},
e^{2i\pi/3}, e^{-4i\pi/3})$ as the end point at $t=1.$  For this
path it is easily checked that $\tilde g(t) f_0 \tilde g(t)^{-1} =
g_1(t) f_0 g_1(t)^{-1}$ with $g_1(t) = \exp( i t Y),$
$$Y=\left(\matrix 0&0&0&0\\  0&0&0&0 \\ 0&0& -2\pi &0 \\ 0&0&0& 2\pi\endmatrix
\right).$$
We can then choose a contraction $g_1(t,r)$ (with $0\leq r\leq 1$) of the loop
$g_1(t)$ to a constant loop. This produces a homotopy  $f(t,r,x)$ connecting
$f(t,x)$ to the constant loop $f_0(x).$

One can check by a straight-forward computation that for the above choices
$$\TR(dh h^{-1})^7 = 14 d\, \TR (df_0 {f_0}^{-1})^5 (dg_1 {g_1}^{-1}).$$
Thus we may apply Stokes' theorem to obtain
$$\int \TR (dh h^{-1})^7 = 14 \int_{S^5} \TR (df_0 {f_0}^{-1})^5 iY.$$
The value of the integral is invariant under the left action $f_0\mapsto af_0$
for any $a\in SU(3).$ In particular, we may choose $a$ such that the matrix
$\int (df_0 {f_0}^{-1})^5$ is of the form $a\cdot \bold{1}+ \lambda,$ where $\lambda$ is
orthogonal with respect to $Y,$ $ \TR \lambda Y=0,$ and $a\in\Bbb C.$ Thus
$$\left(\frac{1}{2\pi}\right)^4\frac{1}{840} \int \TR (dh h^{-1})^7  =
(-2\pi i)\cdot\frac13 \cdot 14\cdot  \left(\frac{1}{2\pi}\right)^4
\frac{1}{840} \int_{S^5} \TR (df_0 {f_0}^{-1})^5
= -n(f_0)/3, $$
where
$$n(f_0) = \left(\frac{-i}{2\pi}\right)^3 \frac{1}{60}
\int_{S^5} \TR (df_0 {f_0}^{-1})^5
$$
is the winding number of the soliton $f_0$ in five dimensions. The factor $1/3$
in the second expression is due to the fact that the trace of the product
$Y \cdot 1_{3\times 3}$ is $-2\pi$ times one third of the trace of the unit
matrix in $SU(3).$

%The generic orbit for the adjoint action in $SU(3)$ is a 4-sphere, so
%geometrically we can think of the $g(t)$'s as a family of rotations on the
%group manifold (the angle $2\pi$ corresponding to the parameter $t=1$).

The loop $t\mapsto f(t,x) = g(t) f_0(x) g(t)^{-1}$ can be viewed as a
rotation on the group manifold $SU(3).$ The above analysis shows that
the action of $\hat\Cal G$ in the fermionic Fock spaces, in the case
of $f(t,\cdot),$ induces a parallel transport around a closed loop
resulting in a phase shift by $e^{-2\pi i n(f_0) /3}.$ When $n(f_0)$ is
not divisible by 3 this is an element of order 3 in $S^1.$

We shall show next that
the interchange of two 5-dimensional solitons in $SU(3)$ of odd winding
numbers leads to the same phase shift in the Fock space.

We assume that $f_1,f_2$ are functions on $S^5$ with values in $SU(3)$
such that the set of points $x\in S^5$ such that both $f_1(x),f_2(x)
\neq 1$ is empty. Let us also assume that $n(f_1) = n(f_2) =1.$
Let $f_0$ be a fixed soliton with $n(f)=1.$ We can choose paths
$g_k(t,x)$ such that $g_k(0,x)=f_0$ and $g_k(1,x)= f_k(x).$
We want to show  that the parallel transport along $g_1(t) g_2(t)$ (at the
end point we obtain the quantum state $q(f_1)q(f_2)\psi$ corresponding to
an initial state $\psi$) is
$e^{2\pi i/3}$ times the parallel transport along $g_2(t) g_1(t)$ in
the Fock bundle (this defines the quantum state $q(f_2) q(f_1)\psi$).
Because the parallel transport depends only on the
homotopy class of the path, with end points fixed, we need to show that
the loop $h(t)=g_1(t) g_2(t) g_1(t)^{-1} g_2(t)^{-1}$ represents of nontrivial
class in $\pi_6(SU(3))$ of degree 3.

Moving around the initial point $f_0$ of the paths $g_k$ does not change
the homotopy class of the loop and therefore we may take $f_0=f_1,$ for
the sake of simplicity. Moreover, we may assume that $f_0$ is a soliton
concentrated in a small neighborhood of the unit element in $SU(3).$
As a consequence, we can take $g_1(t,x)= f_1(x)$
for all $t$ and the homotopy class of $h(t)$ is equal to the class of
$f_1  g_2(t) f_1^{-1} g_2(t)^{-1}.$

Next we note that we can freely move the end point of $g_2(t)$ at $t=1$
so long it does not overlap with the soliton $f_1(x)=f_0(x)$ near the unit
element. Thus we may choose
$g_2(t,x)= a(t)f_1(x)$ where $a(t)$ is path in $SU(3)$ connecting the unit
element $a(0)$ to the element $z$ of degree 3 in the center. Thus after
these deformations,
$$h(t,x)= f_1(x) a(t) f_1(x)^{-1} a(t)^{-1}.$$
It follows that the path $h(t, \cdot)$ is, except for the first constant
factor $f_1,$ a rotation of an (anti)soliton $f_1(x)^{-1}$ in the group
manifold of $SU(3)$ as discussed above. Because the curvature  in the
group extension is zero, the parallel transport depends only on the endpoints
and the homotopy class of the path and consequently the parallel transport
is given by the same phase shift $e^{-2\pi i n(f_1^{-1})/3}=e^{2\pi i/3}$ as
in the case of a rotation.

The homotopy group $\pi_6(SU(3))$ has a generator which is of order 6, so
the element of order 3 which we have constructed is the square of the
generator. However, the square root of our homotopy class is not
described by a  simple explicite formula.

One can perform a similar analysis for other groups and dimensions.  For example,
$\pi_{2n-1}(SU(n))=\Bbb Z$ and there are $SU(n)$ solitons in $2n-1$ dimensions;
$\pi_{2n}(SU(n))=\Bbb Z_{n!}$
and from this one can reduce that there are soliton configurations with
a factor $q\neq 1$ in the commutation relations $\hat f_1\cdot \hat f_2 +q
\hat f_2 \cdot \hat f_1 =0$ such that $q^k=1$ when $k$ divides
$n!.$ Actually, this class of homotopy groups has been discussed in the
context of a space-time formulation of global gauge anomalies, [EN].

\vskip 0.3in
\bf References \rm

\vskip 0.3in
[CMM1] A. Carey, J. Mickelsson, and M. Murray: Index theory, gerbes, and
hamiltonian quantization. Commun. Math. Phys. \bf 183, \rm 707 (1997)

[CMM2] A. Carey, J. Mickelsson, and M. Murray: Bundle gerbes applied to quantum
field theory. hep-th//9711133

[DP] C.T.J. Dodson and Philip E. Parker: \it A  User's Guide to Algebraic
Topology. \rm Kluwer Academic Publishers, Dordrecht-Boston-London (1997)

[EN] S. Elitzur and V.P. Nair:  Nonperturbative anomalies in higher
dimensions. Nucl. Phys. \bf B243, \rm 205 (1984)

[FR] D. Finkelstein and J. Rubinstein: Connection between spin, statistics, and
kinks. J. Math. Phys. \bf 9, \rm 1762 (1968)

[F-Sh] L. Faddeev, S. Shatashvili: Algebraic and hamiltonian methods in the
theory of nonabelian anomalies. Theoret. Math. Phys. \bf 60, \rm 770 (1984)

[M1] J. Mickelsson: \it Current Algebras and Groups. \rm Plenum Press, London -
New York (1989)

[M2] J. Mickelsson:  Chiral anomalies in even and odd dimensions. Commun.
Math. Phys. \bf 97, \rm 361 (1985)

[NA] P.Nelson and L. Alvarez-Gaume: Hamilton interpretation of anomalies.
Commun. Math. Phys. \bf 99, \rm 103 (1985)

[W] E. Witten: Current algebra, baryons, and quark confinement. Nucl. Phys.
\bf B223, \rm 433 (1983)

\enddocument